# Measuring ionizing radiation in the atmosphere with a new balloon-borne detector

K. L. Aplin[1], A. A. Briggs[1], R. G. Harrison[2] and G. J. Marlton[2]

[1] Physics Department, University of Oxford, Denys Wilkinson Building, Keble Road, Oxford, OX1 3RH UK

[2] Department of Meteorology, University of Reading, PO Box 243, Earley Gate, Reading RG6 6BB UK

Corresponding author: Karen Aplin (karen.aplin@physics.ox.ac.uk)

**Key Points:**

- New detector using CsI(Tl) scintillator with photodiode readout developed and tested on meteorological radiosonde
- Particle energy identification is possible from pulse height
- Detector is sufficiently inexpensive to be disposable and deployed routinely

**Abstract**

Increasing interest in energetic particle effects on weather and climate has motivated development of a miniature scintillator-based detector intended for deployment on meteorological radiosondes or unmanned airborne vehicles. The detector was calibrated with laboratory gamma sources up to 1.3 MeV, and known gamma peaks from natural radioactivity of up to 2.6 MeV. The specifications of our device in combination with the performance of similar devices suggest that it will respond to up to 17 MeV gamma rays. Laboratory tests show the detector can measure muons at the surface, and it is also expected to respond to other ionizing radiation including, for example, protons, electrons (>100 keV) and energetic helium nuclei from cosmic rays or during space weather events. Its estimated counting error is ±10%. Recent tests, when the detector was integrated with a meteorological radiosonde system, and carried on a balloon to ~25 km altitude, identified the transition region between energetic particles near the surface, which are dominated by terrestrial gamma emissions, to higher-energy particles in the free troposphere.

**1 Introduction**

Earth's atmosphere is constantly bombarded by energetic particles, mainly galactic cosmic rays (GCR), plus occasional space weather events introducing additional particles to the stratosphere and troposphere [*Bazilevskaya*, 2005]. Despite this continual impingement, the effects of energetic particles in the lower atmosphere (here taken to mean the boundary layer, troposphere and lower stratosphere) are still poorly understood. There are many mechanisms by which weather and climate could potentially be affected, discussed fully by *Mironova et al.* [2015], but most effects of energetic particles in the lower atmosphere are related to their ability to ionize the air. The ions created may play a direct role in formation of aerosol particles [e.g. *Duplissy et al.*, 2010], or accumulate on cloud edges, affecting the microphysics [*Harrison et al.*, 2015]. Atmospheric ions can directly absorb infra-red radiation [*Aplin and McPheat*, 2005; *Aplin and Lockwood*, 2013], and high energy particles are also suspected of influencing lightning rates [*Scott et al.*, 2014]. This paper describes a new technique for measuring and identifying energetic ionizing radiation in the atmosphere.

Ionization near the land surface is principally gamma and alpha radiation from rocks containing uranium and thorium and their decay series. Above the continental boundary layer (the lowest 1-2 km of the atmosphere), and over the sea, GCR are the dominant source of ionization. Occasionally, during space weather events, there are additional ionization bursts from solar energetic particles (SEP), but only the most energetic particles, of ~1GeV or more, are thought to reach the troposphere, whereas particles of energy greater than ~100MeV can access the stratosphere [*Bazilevskaya*, 2005]. For space weather events with a suitably energetic particle spectrum, the ionization rate can be significantly enhanced in the lower atmosphere, even down to the surface; this is known as a Ground Level Enhancement or a Ground Level Event (GLE) [*Bazilevskaya*, 2005]. Despite their transience, space weather events can significantly modulate the local atmospheric electrical environment [*Aplin and Harrison,* 2013; *Nicoll and Harrison,* 2014] and may therefore affect meteorological processes through the mechanisms described above, in addition to other space weather hazards.

Measurements of atmospheric ionizing radiation away from the surface are long-established [e.g. *de Angelis*, 2014] and famously led to the original discovery of cosmic rays by Victor Hess from a balloon flight in 1912. Initially, balloon-borne instruments measured ion currents from ionization chambers, because this was the only practical way to detect energetic particles. Some important measurements of the atmospheric ionization profile continued to be obtained using this technique until the 1970s [*Neher*, 1971; *Harrison and Bennett,* 2007]. However, rapid progress in particle physics following the discovery of GCR meant that the Geiger counter became the preferred instrument for atmospheric energetic particle measurements, partly because ionization chambers were insensitive to the lower energy part of the GCR spectrum [*Bazilevskaya et al.*, 2008]. A regular series of balloon flights carrying Geiger counters to measure energetic particle fluxes started at the Russian Lebedev Physical Institute in 1957, and continues to the present day [e.g. *Stozkhov et al.*, 2009]. This data series remains the definitive source of within-atmosphere energetic particle profiles. Recently, however, ongoing research into the atmospheric effects of energetic particles has motivated additional measurements. The available data is mainly from limited balloon ascents, usually on special campaigns carrying expensive payloads for recovery after descent [*Mertens*, 2016], and surface GCR measurements from the worldwide network of neutron monitors. It has become clear that one-off campaigns and neutron monitor data do not adequately represent the spatial and temporal variability of measured ionization profiles [*Aplin et al*., 2005; *Harrison et al.*, 2014; *Makhmutov et al*., 2014], particularly in the weather-forming region of the atmosphere. Sophisticated models of atmospheric ionization now exist, but rely on standard thermodynamic profiles [*Usoskin and Kovaltsov*, 2006]. Recent measurements have also suggested interesting effects, such as the penetration of lower energy particles further into the atmosphere than was initially thought possible [*Nicoll and Harrison*, 2014], and an influence on lightning [*Scott et al*., 2014; *Owens et al.,* 2015]. There is therefore an ongoing need for more comprehensive measurements of energetic particles in the atmosphere, ideally made in combination with atmospheric properties. Regular measurements such as those exploiting the existing radiosonde ascents of meteorological services will require a miniaturized instrument. Although Geiger counters are the most established technology for atmospheric particle or ionization profiles, these require a high voltage supply, and their use can be subject to supply constraints. They also do not provide energy discrimination, hence the origin of the events measured cannot be readily determined.

In this paper a new instrument is presented that is low-current (20mA), low-voltage (12V) and low-mass (30g), with energy discrimination. It is inexpensive enough (a few hundred GBP) to be deployed on routine meteorological radiosonde launches to obtain combined atmospheric and particle data. Section 2 describes the instrument and its calibration, and Section 3 presents the results of two test flights measuring atmospheric energetic particles.

## 2 Instrument Description and Laboratory Tests

The sensor used is a 1x1x0.8 $cm^3$ caesium iodide thallium activated scintillator (CsI:Tl) with an integrated 1$cm^2$ silicon PIN (*p*-i-*n* junction) photodiode (First Sensor X100-7THD), operated under reverse bias at 12V. CsI:Tl scintillators are more mechanically robust with a greater intrinsic efficiency per unit volume than similar scintillators such as NaI:Tl, and the wavelength of their optical output is especially optimal for photodiode detection [*Fioretto et al*., 2000; *Saint-Gobain,* 2016], making them ideal for this application. The principle of operation is that ionizing radiation creates a flash of light in the scintillator, which is detected by the photodiode and converted into a charge pulse, proportional to the energy of the incoming particle. (Neutrons are an important component of the secondary cosmic radiation but do not contribute to atmospheric ionization. The scintillator used is ideal for detecting atmospheric ionization because the caesium, iodine and thallium it contains all have a small cross-section for neutron interaction.) The associated current pulses are converted to voltage pulses by a fast rise time current to voltage convertor [*Aplin and Harrison,* 2000] using a LMP7721 opamp. After further amplification, the voltage pulses are compared with a slowly responding mean voltage level (the trigger threshold) using a Schmitt trigger. When the Schmitt trigger detects an event, an interrupt signal is generated at the microcontroller and the pulse height and background reference level are measured, from which a timestamp for each pulse, the pulse height and reference voltage are provided as ascii serial data. The sensor can operate as a standalone unit or, with suitable interfacing, as an accessory for a meteorological radiosonde [e.g. *Harrison et al.*, 2012] or remotely piloted vehicle. Data can be output via USB, Bluetooth, or on request over the serial bus. The instrument operation is summarized in Figure 1. Unlike a conventional Geiger counter, this detector can be run from a low bias voltage (12V) at a low current (20mA) and also has the important advantage of being able to resolve particle energy, which allows identification of different types of ionizing radiation.

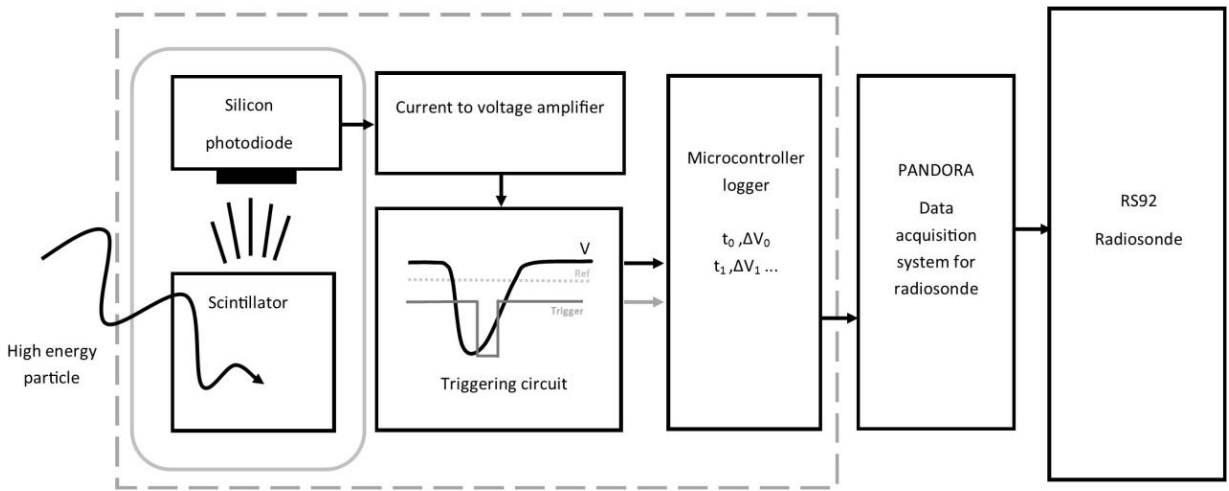

Figure 1. Block diagram of the instrument operation showing the signal flow. In this case the output has been attached to a radiosonde.

Near the continental surface, atmospheric ionization has a typical volumetric production rate of ~10 cm$^{-3}$s$^{-1}$ [e.g. *Chen et al.,* 2016]. This rate is dominated by gamma rays originating from the decay chains of naturally radioactive uranium-238 and thoron-220 in the ground, with a ~20% contribution from GCR [e.g. *Chen et al.,* 2016; *Minty*, 1992]. (Alpha and beta particles contribute ~10% but cannot enter the detector, and will not be considered further.) The energy deposited by gamma rays in matter is well-understood and originates from several physical processes, giving a well characterized shape for all gamma ray spectra [e.g. *Knoll*, 2010]. The maximum energy peak in a gamma ray spectrum results from the "photopeak" when all the particle energy is deposited in the detector through photoelectron absorption. Other processes (Compton scattering and pair production) lead to further signals at lower energies, with the detailed response depending on the detector size and geometry.

Away from the surface, all the ionization is from GCR (both primaries and secondaries) and, episodically, space weather events, both of which are generally higher energy than the terrestrial gamma ray particles of ~500 keV – 3 MeV. Of the higher energy particles routinely reaching the lower atmosphere, muons dominate ionization near the surface, with a contribution from the "electromagnetic cascade" of electrons, positrons and photons, which becomes the major source of ionization higher up in the atmosphere [*Usoskin and Kovaltsov*, 2006].

Tropospheric muons are sufficiently high energy (~4GeV) to be considered "minimum ionizing particles", i.e. they pass through the scintillator with minimum loss of energy from ionization, and are likely to produce a very small signal [e.g. *Nakamura et al.,* 2011]. Other atmospheric particles from space weather and cosmic rays are also expected to be detectable. Photons will produce gamma ray-like spectra, and other ionizing radiation, if sufficiently energetic (>100 keV for electrons and positrons, based on their ability to penetrate the detector housing) will interact with the scintillator to produce an optical pulse proportional to their energy. (Alpha particles are not anticipated to be able to penetrate the detector housing.) The scintillator-photodiode combination used is specified to respond to 1-10 MeV, but an almost identical detector (1 cm$^3$ with photodiode readout) responded to energetic particles up to 25 MeV [*Viesti et al.,* 1986]. In practice, the particle energies the detector is sensitive to are limited at the lower end by noise, and at the upper end by the compromises between gain and resolution in the amplification circuitry. In the next section the efficiency and energy response of the detector are discussed.

### 2.1 Laboratory calibration

This section describes the response of the detector as measured in the laboratory, both with radioactive sources, and natural background radioactivity.

2.1.1 Background count rate

Several background tests were run in different laboratories within the Oxford University Physics Department between March and December 2016. Mean count intervals for these separate tests of 17-48 hours varied from 36.5s between counts to 54s between counts. Given that this detector has an area of 1 cm$^2$, these background count rates are slightly larger than the textbook GCR flux of 1 cm$^{-2}$ min$^{-1}$ [*Nakamura et al*., 2011], however this is expected, as there will be some contribution from gamma rays. The intervals between the counts fitted an exponential distribution, as expected for all types of radioactive decay, indicating that the measured background count rate is generated from a combination of GCR and gamma rays.

In a separate experiment, the pulse counting electronics was tested by connecting the detector to a digital storage oscilloscope (LeCroy Waverunner 204 MX1 2GHz) capable of recording each pulse sensed by the scintillator. The oscilloscope trigger was set to be sensitive to every possible radioactive pulse, in addition to many "false" triggers from noise. The shape of each pulse was sampled by recording data on the oscilloscope's hard drive and the recorded pulses were subsequently inspected to visually determine which were associated with radioactivity, which produces a characteristic shape of response [*Knoll*, 2010] compared to noise. In the flight instrument, as in any energetic particle detector, there is necessarily a compromise between detecting false positives from noise, or the possibility of missing lower energy particles that are close to the noise threshold. In this experiment, no false positives were recorded by the new detector, but approximately 10% of valid pulses from energetic particles were missed due the limitation in our trigger design, providing a 10% uncertainty on measured pulse rates.

2.1.2 Energy calibration

Energy calibration is carried out with well-characterised radioactive sources emitting gammas at different energies and activity levels, Table 1. Preliminary tests indicated that the scintillator detector did not respond to a 136 keV Co-57 X ray source, though a signal was detected in a version with a PIN photodiode alone (as is well-established [*e.g. Freck and Wakefield*, 1962]). A strong motivation for choice of the scintillator and PIN photodiode combination is its preferential response to more energetic particles, the main contributors to atmospheric ionization, which makes it more relevant for this application than a detector responding to lower-energy particles.

| **Radioisotope** | **Gamma energy emitted (keV) (photons per 100 disintegrations)** | **Activity of source on 24/3/16 (Bq)** |
|---|---|---|
| Caesium-137 | 662 (85) | 66325 |
| Cobalt-60 | 1173 (100), 1332 (100) | 9418 |
| Sodium-22 | 511 (181), 1275 (100) | 1839 |

Table 1. Summary of laboratory sources used for calibration. Gamma energies are taken from *Bé et al.*, [2010]

Energy calibration was carried out by running the detector when irradiated by gamma sources, figure 2. The tests with sources were run for 3 hours each, and the background test for 17 hours, in a basement laboratory at Oxford University Physics Department on 23$^{rd}$-24$^{th}$ March 2016. Most background gammas were excluded from the source tests with lead shielding. Figure 2(a) shows histograms of the raw data. The characteristic shape of the histograms is related to the detector's trigger circuitry, which excludes pulses $< \sim 150$ mV, and the scintillator's quantum efficiency, which is 1 for particles $< \sim 200$ keV and decreases steeply with energy. Figure 2(b) shows the response of the sources with the background signal subtracted. By assuming that the major peaks seen for each source are caused by the photopeak, the voltage at which local peaks in the distribution occur are associated with the gamma emissions from each source (Table 1).

The Cs-137 661keV photopeak is assumed to be the peak in the centre of figure 2(b), at about 275mV. The broader Na-22 peak at ~250mV can then be linked to its 511keV gamma, and the co-located peaks for Na-22 and Co-60 at about 375mV are taken to be ~1 MeV gammas (see Table 1 for the gamma energies emitted by each source). Since no clear difference can be seen between these three peaks, they are assumed to be unresolved by the detector. If the 375mV peak is caused by these ~1 MeV gammas, which have mean energy $1255 \pm 40$ keV, then three points are available for calibration. These three points lie on a line, to which a linear fit between source energy and pulse height has an adjusted $R^2$ of 0.9935. This three-point fit encourages the view that the peaks used in the plot are valid photopeaks; if they were not, then a good linear fit could only be obtained by chance, for which the probability, *p*, is <0.04, generally regarded as low.

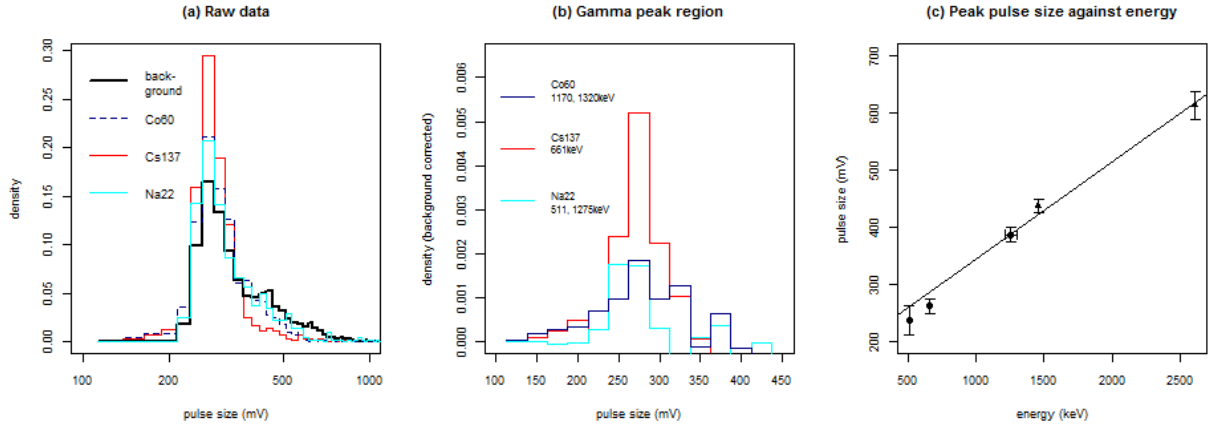

Figure 2. Laboratory energy calibration measurements (a) raw data, showing a histogram of pulse sizes for background and three sources (b) histogram of the source peak region with background subtracted (c) pulse height against energy, obtained by linking the voltage at which local peaks in the distributions occur with each source's photopeak. The Co-60 peaks at 1170 and 1320 keV and the Na-22 peak at 1275 keV are assumed to be unresolved by the detector and are plotted as a mean of 1255 ± 40 keV. The two higher-energy peaks are from the natural radioactive isotopes K-40 and Tl-208, indicated by triangular data points (data points obtained from laboratory sources are circular). Error bars are based on the estimated error in pulse height measurement.

Additional calibration points can be obtained from background natural radioactivity, which has gamma ray signals at higher energies than those from the available laboratory sources, including the well-known peak from K-40 at 1460 keV and Tl-208 at 2610 keV [e.g. *Minty*, 1992]. These can be seen in figure 2(a) above the source signals at ~475 mV and ~600mV respectively. The voltages corresponding to the peaks from the three sources and background natural radioactivity have been plotted in figure 2(c). The calibration between energy *E* in keV and pulse size *P* in mV is then obtained from a five point least squares fit (adjusted $R^2$ = 0.9881, $p < 0.0003$), given by:

$$P = (0.177 \pm 0.009)E + (160 \pm 15) \quad \text{(eq 1)}.$$

This calibration is used to calculate the energy of ionizing radiation detected later in the paper.

In a separate experiment carried out by providing random triggers obeying Poisson statistics to the sampling microcontroller, the mean error in each pulse height measurement is ~10mV with a standard deviation $\sigma$ of 29mV. According to gamma detector theory [e.g. *Knoll*, 2010], the resolution of each peak is given by $2.35\sigma = 69$ mV, which, from the calibration above, would be ~400 keV. This is consistent with the assumption made above that the peaks between 1100 and 1300 keV cannot be distinguished, although the identification of the K-40 peak at 1460 keV and separation of the Cs-137 and Na-22 peaks at 662 and 511 keV indicates that the actual resolution may be ~100keV, better than the value determined in the false triggering experiment.

The measured response is comparable with the quoted noise level of 70 keV for detectors of similar size and geometry [*Saint-Gobain,* 2008].

Although the calibration experiments are only carried out up to a maximum energy of 2.6 MeV, as identical detectors respond up to 25 MeV [*Viesti et al.*, 1986] it is reasonable to extrapolate our calibration range up to the maximum measurable pulse height, of 3.2 V, which corresponds to a gamma energy of ~17 MeV.

2.1.3 Detector efficiency

In principle, it is possible to calibrate any particle detector's response to a known source with a known activity level. Particle detectors have a quantum efficiency based on how many gamma interactions occur within the detector. This quantum efficiency is provided by the manufacturer in their data sheet, and for our detector, falls rapidly with energies >200 keV as discussed in section 2.1.2, and levels off at ~ 7 % for energies > 5 MeV.

There is a further efficiency effect based on the types of gamma ray interaction occurring in the detector. Section 2.1.1 above showed that our detector responds well in the photopeak region for typical gamma sources i.e. in the part of the spectrum where all the energy of incoming gammas is deposited in the scintillator. According to the intercept of the calibration equation (eq 1), the detector should be able to respond to the entire gamma ray spectrum down to a few keV. It seems likely that the lower-energy interactions, when not all the energy of a gamma is deposited in the detector, are close to the threshold below which pulses cannot be unambiguously identified over the noise, so that the full gamma ray spectrum cannot be measured. This will result in a loss of efficiency in the detector in the sense that the number of counts recorded is not equivalent to the number of counts expected from a given radioactive source. This "peak to total" ratio is not straightforward to determine, but is estimated to be 10% for 1 MeV gammas for a 1 $cm^3$ Cs:I(Tl) scintillator [*Saint-Gobain,* 2008]. This was consistent with results from laboratory tests with radioactive sources at a fixed distance from the detector, where the efficiency in response to 1 MeV particles from Co-60 was 5.8±0.6%.

To summarise the detector's measured response,

- Background count rates and the count interval distribution are consistent with an instrument that responds to gamma rays and GCR muons. The counting error is ±10%.
- The instrument can detect the photopeaks of laboratory radioactive sources, and this permits energy calibration for gammas (Eq 1) from ~500 keV – 2.6 MeV.
- Gammas up to 17 MeV are expected to be detectable.

**3 Test Flights**

The device was tested on two radiosonde flights on 16th August and 6th October 2016, from Reading University Atmospheric Observatory (51.46° N, 0.95° W), UK. For both flights the ionization sensor was interfaced to a Vaisala RS92 radiosonde through the Programmable ANd Digital Operational Radiosonde Accessory (PANDORA) [*Harrison et al.,* 2012], figure 1. This allows data from the ionization sensor to be relayed back to the ground station in real time with height and GPS position data from the radiosonde, as well as meteorological data such as pressure, temperature and humidity. As the bandwidth of the radiosonde's radio link for extra sensor data is 64 bits / second, relaying a full event list is not possible, especially in atmospheric regions with substantial count rates. Thus, the microcontroller on the radiation detector board measures the size of any pulses by differencing the pulse height from the reference level and bins each count into a histogram. The PANDORA sends a regular request for data to the ionization sensor's microcontroller, in response to which the histogram data is returned for transmission over the radio link, as well as the raw voltage values for the reference level, pulse height for the last pulse before the data request, and the total number of pulses received since the last request. Transmitting the detailed information for the last pulse received is a useful indicator of the raw voltages measured, and can also provide some redundancy in case there are problems with the pulse counting or histogram data. The histogram bins and associated data is buffered in the PANDORA board microcontroller and relayed over the radio link; due to the bandwidth this transfer takes about 4 seconds.

3.1 Summary of test flights

The circumstances of each launch are summarized in Table 2. The detectors were switched on for a short period before the launch to obtain a background count rate and spectrum, with the background count rates shown to be comparable between the launch site and a laboratory in the Reading University Meteorology Building, approximately 250m south east of the Observatory where the launches took place (Figure 3). The background count rates in an unscreened environment are approximately a factor of three higher than in the laboratory tests; this is consistent with a greater contribution from gamma rays in the open air. If the counts recorded in the background experiment results were mainly from GCR, then the unscreened count rate is consistent with the partitioning of atmospheric ionization, in which the contribution of gamma rays is about 3.5 times greater than GCR (Section 2). Energy histograms during these reference runs are shown in Figure 4 and can be seen to be consistent. In figure 4(a) only 8 counts in total were recorded, but in figures 4(b) and 4(c) the statistics are more reliable, and show that there is little day-to-day variation in the local particle spectrum at the Reading University site, indicating that the counts are mainly from terrestrial gamma radiation, associated with the local geology.

Manuscript submitted to *Space Weather*|  | **Launch 1** | **Launch 2** |
|---|---|---|
| Date and time (UT) of launch | 16$^{th}$ August 2016 14:31 | 6$^{th}$ October 2016 15:06 |
| Time of last communication received and location | 16:53 (8393.4s after launch) 51.65 °N, 0.43 °W | 17:21 (8113.4s after launch), 50.79 °N, 1.22 °W |
| Time, height and pressure of balloon burst | 16:04 (5614.6s after launch), 29.1 km, 14.5 hPa | 16:43 (5836.6s after launch), 29.2 km, 13.3 hPa |
| Seconds after launch, altitude, pressure and temperature of noise onset | 118 s, 638 m, 944.6 hPa, 16.1 °C | 152 s, 968 m, 909.1 hPa, 5.3 °C |
| Median and standard error neutron counts at Kiel during flight | $(10365 \pm 11)$ min$^{-1}$ | $(10499 \pm 17)$ min$^{-1}$ |
| *Daily average space weather conditions* |  |  |
| Sunspot number: | 70 | 38 |
| Kp index: | 2 (quiet) | 2 (quiet) |
| Interplanetary magnetic field: | 4.4 nT | 8.9 nT |
| Solar wind protons from ACE spacecraft: | 4.2 cm$^{-3}$ | 5.5 cm$^{-3}$ |

Table 2. Summary of test flights launched from Reading University Atmospheric Observatory (51.46° N, 0.95° W). Cosmic ray neutrons and space weather conditions at the time of the launches are also indicated, and were obtained, respectively, from the Kiel neutron monitor (54.3° N, 10.10° E) and spaceweather.com.

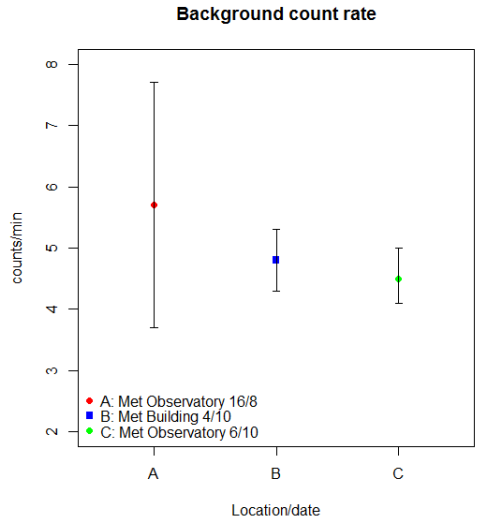

Figure 3. Background count rate before each radiosonde launch. Error bars are determined from the square root of the number of counts in each test.

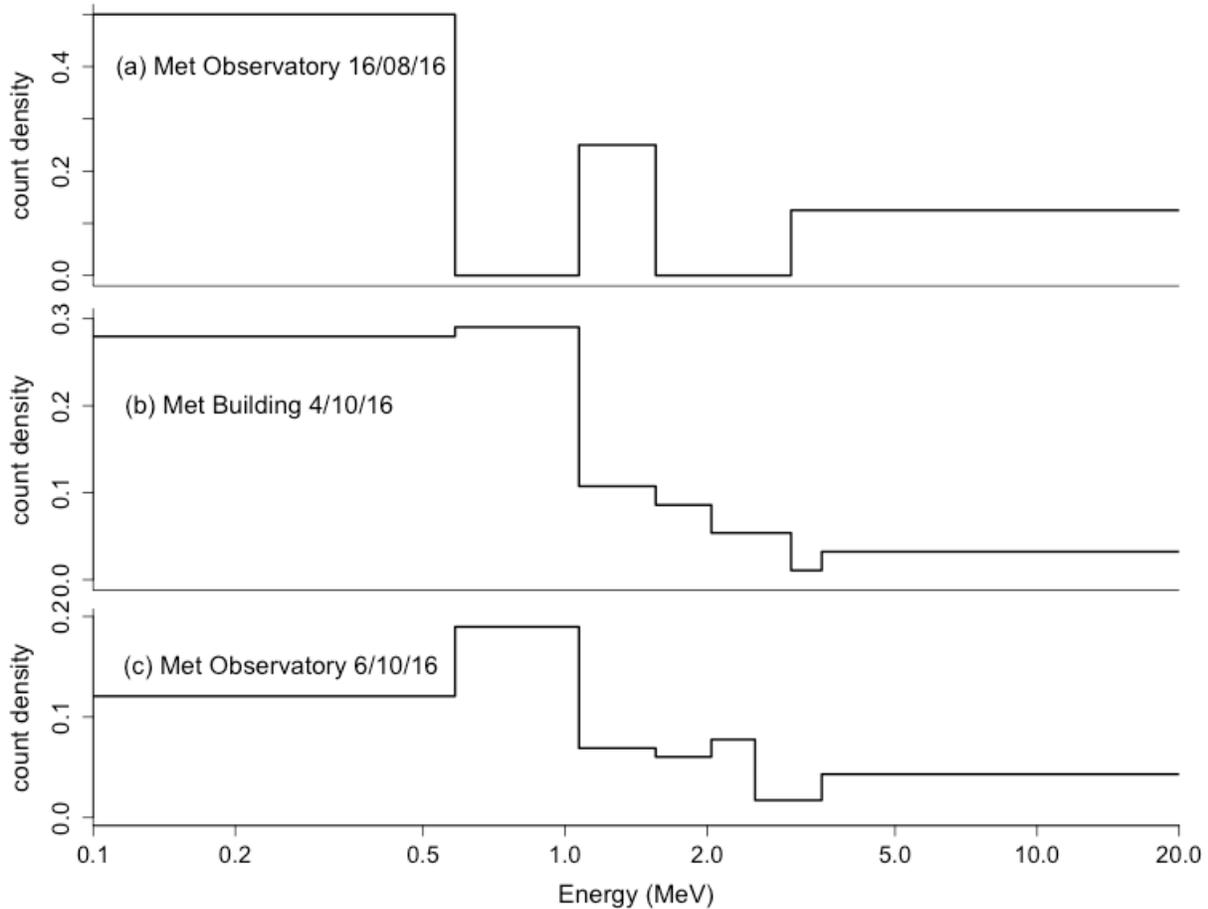

Figure 4. Particle energies measured during pre-launch tests (a) immediately before, and the first part of, the launch on 16[th] August 2016 (b) laboratory test at Reading Meteorology Department on 4[th] October 2016 (c) immediately before, and the first part of the launch on 6[th] October 2016.

A few minutes after the August 2016 launch, the count rates increased dramatically and did not show the expected variation with height during the flight. These extra counts occurred from the pressure level of 900 hPa, consistent with the transition to the high power mode of the transmitter, implemented by the radiosonde manufacturer at about ~50 hPa below the surface pressure [*Vaisala*, 2014] to protect the radiosonde receiver. After this problem arose during the August launch, laboratory tests were carried out with the final assembly in a chamber, in which the pressure was systematically reduced to switch the radiosonde into its high power mode, whilst measuring its radio frequency (RF) emissions. The onset of high power mode was associated with spurious counts caused by noise-induced false triggering. In the October pre-flight tests, the sensitivity to RF noise was removed in the laboratory by inclusion of extra decoupling components in the circuitry, and, although the noise effects were reduced, some erroneous pulses were again generated. Nevertheless, as the reference voltage and pulse height were recorded for one pulse per transmission, it has been possible to examine some of the individual pulses received. Using this remedial technique it is possible to retrieve particle energy, but not count rates.

3.2 Measured particle energies

In the laboratory calibration tests the reference voltage for the trigger stage was a stable 3.221±0.0005 V (where the error is the standard error on the mean) and in the pre-flight testing at Reading on 4$^{th}$ October it was also stable, at 3.221±0.002V. The radio transmitter interference is episodic and induces voltage fluctuations, with larger fluctuations interpreted as triggers by the detector's circuitry. During periods with no false triggers, the detector can still measure energetic particles. (The reference voltage was designed to drift with temperature in the same way as the signal line, so that common mode variations in temperature are experienced, to minimize uncertainty.)

Heights of each of the individual pulses received per transmission (approximately 1 every 14s) were calculated by differencing. Pulses with a "negative" size were rejected as unphysical, but all other pulses are included in the data analysis. In the lowest layer of the atmosphere, the boundary layer, the pulse heights are expected to be typical of terrestrial gamma emissions [*Minty*, 1992] whereas at higher altitudes they will be dominated by higher-energy GCR, and any particles from space weather events. The free tropospheric particle spectrum is therefore expected to be higher energy than the near-surface particle spectrum. To assess this, the data values have been separated into boundary layer and "cosmic" regimes of heights less than 1-2 km and greater than 8 km, where gamma and GCR contributions respectively are expected to be dominant. The boundary layer height was determined from the temperature profile measured in each ascent. The scintillator's output decreases with temperature due to changes in the light-generating transitions within the crystal [*Valentine et al.*, 1993], so a correction was applied to take account of this effect [*Saint-Gobain*, 2016]. The results are summarized in Figure 5. The responses from each flight are similar, with the near-surface gamma energies indistinguishable at 1.6 – 1.7 MeV, consistent with the results reported by *Minty* [1992]. The median cosmic pulse heights were 911 mV and 733 mV for August and October respectively. If the gamma calibration in eq (1) is assumed, the energies involved are 4.2 ± 0.3 MeV for the August flight and 3.2 ± 0.2 MeV for the October flight, therefore it is concluded that the detector has identified a transition between terrestrial gamma rays and GCR. There was no significant difference in the GCR or space weather characteristics on each of the flight days. GCR data were obtained from the Kiel neutron monitor [e.g. *Moraal et al.*, 2000] which has a similar cutoff rigidity (2.36 GV) to Reading (3.64GV). Space weather conditions were quiet for both launches (Table 2).

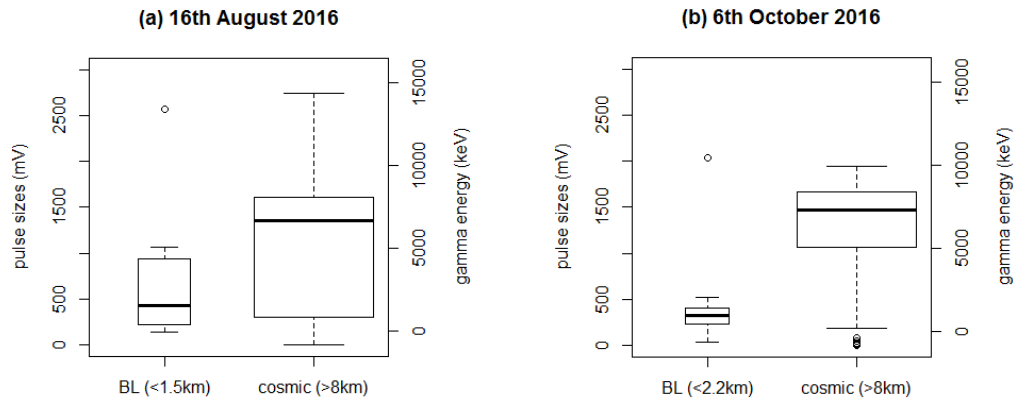

Figure 5. Boxplots of pulse height in mV (left hand axis), calibrated to gamma energy (right hand axis) for the boundary layer (BL) height determined from the measured temperature profiles for the launch, and in the free troposphere for (a) August flight and (b) October flight. The width of each box is scaled according to the square root of the number of points it contains, and the thick line denotes the median pulse size. The whiskers are the first and third quartiles of the data with outliers beyond this range plotted as individual points.

**4 Conclusions**

A new type of detector for atmospheric energetic particle measurements using novel low-cost electronics has been described, characterized in the laboratory and tested on meteorological radiosonde flights. It responds to the photopeak from gamma rays, which allows a simple energy calibration with laboratory sources and natural radioactivity up to ~3 MeV, but, as was discussed in section 2, the detector is expected to measure up to 17 MeV gammas. The instrument can respond to muons, producing an appreciable count rate when covered with a thick layer of lead, however, we are not currently able to distinguish muons from gammas in the air around the detector, as the muons, though highly energetic, are "minimum ionizing particles" and produce a signal of similar magnitude to the background gamma radiation. The detector is also expected to respond to all atmospheric ionizing radiation including primary cosmic ray particles (principally helium nuclei and protons) and electrons with sufficient energy to penetrate the detector housing (> ~100keV). It is not thought to be able to detect neutrons, due to a small neutron cross-section of the materials in the scintillator, and is therefore uniquely sensitive to the majority of the particles responsible for atmospheric ionization.

In test flights launched from Reading, UK in August and October 2016, energy separation of a high-energy tropospheric and a lower-energy near-surface component was obtained. Terrestrial gamma energies were unchanged from August to October, as expected since they are dominated by local geology. The GCR background energy under calm space weather conditions was also similar for each flight. The system can therefore separate different types of particle on the basis of their energy. There is scope for more detailed particle identification based on pulse shape [*Skulsi and Momayezi,* 2001], by further development of our low-cost microcontroller pulse analyser technology. Additional laboratory experiments with careful exclusion of background gammas should identify the muon signal more clearly. Modelling work using, for example, the GEANT-4 package would also improve our ability to separate signals from different types of ionizing radiation in the detector [e.g. *Fioretto et al.*, 2000].

Subsequent flights will implement better screening to reduce the interference problems identified in these tests, and permit assessment of the response to e.g. primary GCR. Measurements during disturbed space weather conditions will also provide an additional opportunity to test the detector. Clearly a laboratory calibration for all the particles and energies expected throughout the entire atmosphere is not practical. It may therefore be helpful to fly the detector as part of an aircraft or high-altitude balloon campaign including a range of sophisticated particle detectors and spectrometers, for comparison [e.g. *Mertens*, 2016]. The detector is versatile enough to be interfaced with remotely piloted vehicles or used in networks of sensors in a wide range of other applications.

## Acknowledgments and Data

This work was funded by the UK Science and Technology Facilities Council (STFC) Impact Accelerator Account and STFC grant ST/K001965/1 *Airborne monitoring of space weather and radioactivity*. Data can be obtained for academic purposes upon request from KLA (karen.aplin@physics.ox.ac.uk). We gratefully acknowledge technical support from A. Baird, P. Hastings, J. Lidgard, K. Long and P. Shrimpton, all at Oxford University Physics Department. Assistance with radiosonde launches was provided by Dr K. Nicoll and Dr M. Airey of Reading University Meteorology Department.